\documentclass[letterpaper, 10 pt, conference]{ieeeconf}  
\usepackage[T1]{fontenc}
\usepackage{amsmath,amsfonts}
\usepackage{algorithmic}
\usepackage{multirow}
\usepackage{algorithm}
\usepackage{array}
\usepackage{textcomp}
\usepackage{stfloats}
\usepackage{url}
\usepackage{verbatim}
\usepackage{graphicx}
\usepackage{color}
\usepackage{booktabs}
\usepackage{xcolor}
\usepackage{adjustbox} 
\usepackage{cite}
\makeatletter
\let\NAT@parse\undefined
\makeatother

\usepackage{hyperref} 
\hypersetup{
    colorlinks=true,
    linkcolor=black,
    citecolor=black, 
    filecolor=magenta,      
    urlcolor=blue,
    }

\IEEEoverridecommandlockouts                              

\title{\LARGE \bf
Towards Robust Multi-UAV Collaboration: MARL with Noise-Resilient Communication and Attention Mechanisms
}

\author{
    Zilin Zhao$^{1}$, Chishui Chen$^{1}$, Haotian Shi$^{2}$, Jiale Chen$^{3}$, Xuanlin Yue$^{4}$, Zhejian Yang$^{5,*}$, and Yang Liu$^{2,*}$%
    \thanks{$^{1}$ College of Software Engineering, Jilin University, Changchun, China (e-mail: zhaozl9921@mails.jlu.edu.cn; chencs5522@mails.jlu.edu.cn).}%
    \thanks{$^{2}$ College of Instrumentation and Electrical Engineering, Jilin University, Changchun, China (e-mail: shiht9921@mails.jlu.edu.cn; liu\_yang@jlu.edu.cn).}%
    \thanks{$^{3}$ School of Communication Engineering, Jilin University, Changchun, China (e-mail: cjl2022@mails.jlu.edu.cn).}%
    \thanks{$^{4}$ School of Electrical Engineering, China University of Mining and Technology, Xuzhou, China (e-mail: 202207020110@stumail.xsyu.edu.cn).}%
    \thanks{$^{5}$ School of Artificial Intelligence, Jilin University, Changchun, China (e-mail: zjyang22@mails.jlu.edu.cn).}%
    \thanks{$^{*}$ Corresponding authors: Zhejian Yang and Yang Liu.}%
}

\begin{document}

\maketitle
\thispagestyle{empty}
\pagestyle{empty}

\begin{abstract}
    Efficient path planning for unmanned aerial vehicles (UAVs) is crucial in remote sensing and information collection. As task scales expand, the cooperative deployment of multiple UAVs significantly improves information collection efficiency. However, collaborative communication and decision-making for multiple UAVs remain major challenges in path planning, especially in noisy environments. To efficiently accomplish complex information collection tasks in 3D space and address robust communication issues, we propose a multi-agent reinforcement learning (MARL) framework for UAV path planning based on the Counterfactual Multi-Agent Policy Gradients (COMA) algorithm. The framework incorporates attention mechanism-based UAV communication protocol and training-deployment system, significantly improving communication robustness and individual decision-making capabilities in noisy conditions. Experiments conducted on both synthetic and real-world datasets demonstrate that our method outperforms existing algorithms in terms of path planning efficiency and robustness, especially in noisy environments, achieving a 78\% improvement in entropy reduction. The supplementary materials are available at \url{https://github.com/zilin-zhao/iros25-supp}.
\end{abstract}

\section{INTRODUCTION}

In recent years, UAV technology has made significant advancements, with its applications expanding across various domains \cite{uav-applications-review}. These developments have provided efficient and cost-effective solutions for numerous industries. Due to their high flexibility, high mobility, and low deployment costs, UAVs are increasingly employed for monitoring complex terrains \cite{uav-monitor-IPP-popovic, uav-monitor-2}. Moreover, in large-scale terrain monitoring tasks, deploying multiple UAVs to explore target areas and establish communication systems for information exchange can significantly enhance exploration efficiency and reduce overall costs \cite{multi-uav-path-planning-1, martin-2}. However, a major challenge in this context lies in designing robust communication protocols \cite{multi-uav-communication-1, siegwart-1-cooperative-IPP} and developing effective decision-making strategies for multiple agents \cite{multi-uav-path-planning-2}. 

The informative path planning (IPP) problem is a critical challenge in such scenarios\cite{uav-monitor-IPP-popovic}. IPP focuses on planning information-rich paths for each agent while considering constrained resource budgets (i.e., time, energy, etc.) to achieve efficient collaborative data collection. Traditional algorithms for IPP problem have been extensively studied \cite{uav-path-planning-Z}, such as non-adaptive methods \cite{bahnemann2017decentralized, non-adaptive-IPP}, which rely on static and pre-defined paths or assume uniform target distributions. However, real-world scenarios are often far more complex. Some studies have proposed adaptive algorithms \cite{hollinger2013active, blanchard2022informative}, enabling UAVs to make dynamic decisions in real time, but as the planning horizon expands, these algorithms suffer from exponentially increasing execution times due to the need to evaluate a vast number of candidate paths, rendering them infeasible for time-sensitive tasks. Meanwhile, GNN-based IPP methods \cite{iros-10, iros-24} have been studied. 
Although these methods can address the execution time issue, they still rely on supervised imitation learning and require expert demonstrations for training.

\begin{figure}[t]
    \centering
    \includegraphics[width=\linewidth]{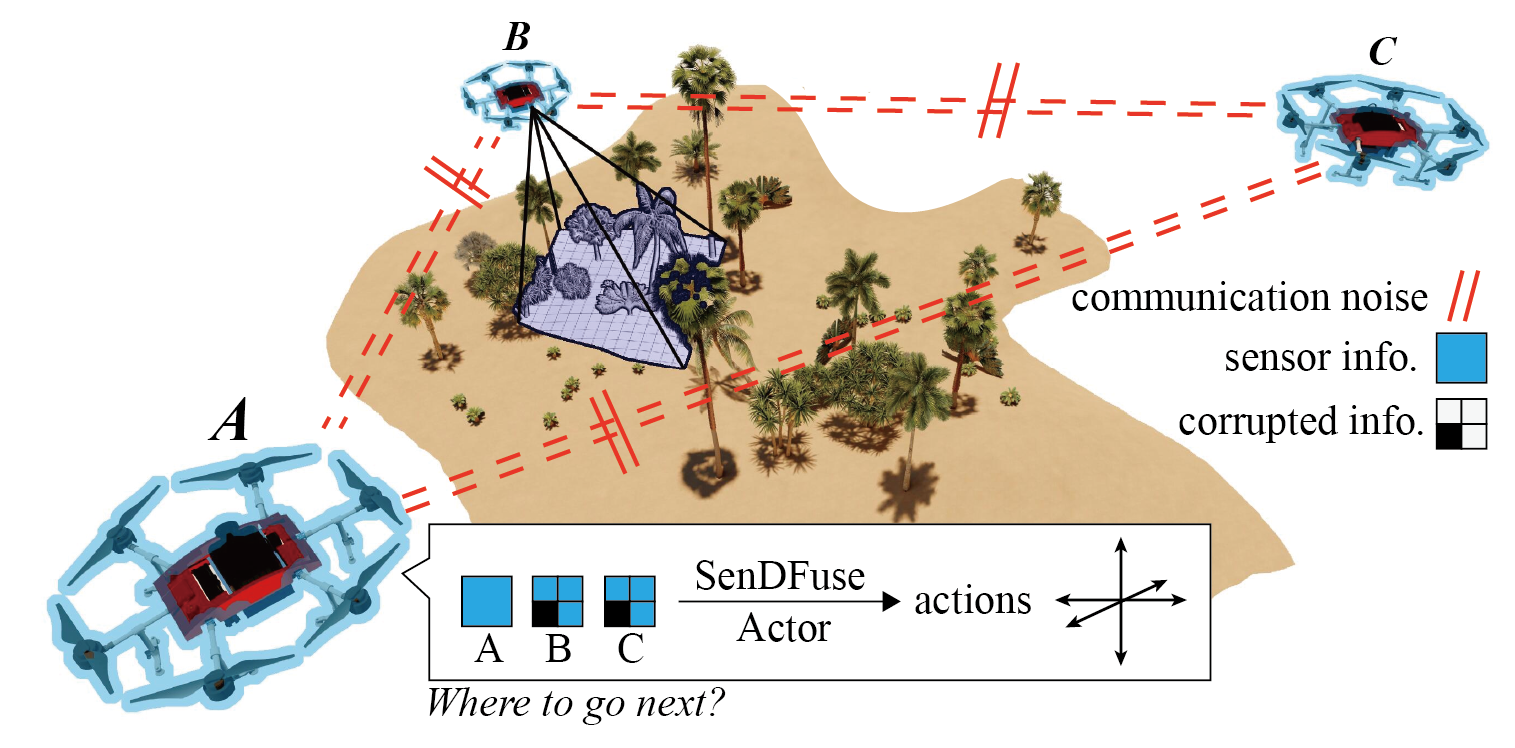}
    \caption{Multiple drones collaborate to scan an area of interest. They communicate in a noisy environment, exchange data, perform sensor fusion with denoising, and  plan paths independently. By working together, they enhance the understanding of the region. }
    \label{fig:intro}
\end{figure}

Reinforcement learning (RL) methods have emerged as a promising approach for effective online decision-making in robotics \cite{RL-robot-1, siegwart-2-RL, gui-3-RL}. By simulating interactions between UAVs and their environment, RL enables UAVs to learn and adapt in dynamic scenarios, progressively improving their path-planning strategies. In the context of the IPP problem, the advantage of RL methods lies in shifting the computational burden of decision-making from the deployment to the training \cite{leanring-based-methods-for-ipp}: with proper training, agents can make high-quality decisions during deployment with constant-time complexity. Furthermore, RL methods do not require supervised training or imitation of artificially defined expert targets. Therefore, Chen et al.\cite{iros-14} and Tan et al. \cite{gui-1-ir2} studied RL for IPP scenarios in 2D spaces, while other reachers' works\cite{iros-18, RL-IPP-1, RL-IPP-4, popo1, CAtNIPP} investigated single-agent RL in 3D environments. In work by Zeng et al.\cite{9811800}, single-agent autonomous exploration for acquiring spatial information about crops in 3D agricultural environments was studied, and Bartolomei et al.\cite{9635893} explored UAV scanning of regions of interest for semantic segmentation, but these works did not address multi-UAV scenarios. Thus, recent works\cite{multi-uav-path-planning-2, iros-12} have studied 3D but fixed-altitude IPP problems. Westheider et al.\cite{RL-IPP-2-baseline} and Wang et al. \cite{gui-2} focusd on solving the credit assignment problem in the multi-agent IPP problem. Between them, the former brought the MARL-IPP problem into full-3D environment for the first time. However, to the best of our knowledge, no existing research addresses the design of multi-UAV communication protocols in this context, particularly under conditions involving sensor fusion and communication noise. Moreover, no prior work has explored the integration of composite attention mechanisms within this framework.

To address these gaps, we propose a novel IPP-MARL framework for full-3D environments, focusing sensor fusion, communication denoising, and attention mechanisms in training and deployment phases. Figure \ref{fig:intro} provides a visual representation of our work. The main contributions of our work are as follows: 
\begin{enumerate}
\item We propose a novel MARL framework that focuses on multi-UAV sensor fusion and denoising, and incorporating attention mechanisms into both the decision-making and training stages of MARL.

\item We develop a multi-sensor denoising and fusion network, termed the SenDFuse Network, to establish a robust communication protocol for multi-UAV systems.


\item We utilize the COMA algorithm \cite{COMA} to provide a credit assignment foundation for the proposed framework, and we validate its potential for engineering applications through training and deployment experiments in both synthetic and real-world data.
\end{enumerate}

Our paper is structured into five sections. The current section provides a brief overview of the research landscape and surveys related works. In Section \ref{problem-statement}, we present a formal definition of the research problem. Section \ref{method} details our proposed method through four subsections: the overall framework, the SenDFuse Network, Actor and Critic networks, and the training algorithm. Comprehensive experimental results using both synthetic and real-world data are presented in Section \ref{experiment}, and Section \ref{conclusion} concludes the paper.

\section{Preliminary}
\label{problem-statement}
Our work focuses on a multi-UAV cooperative navigation task, aiming to enable the system to generate high-information-density paths in noisy environments and constrained communication budgets. The goal of the system to enhance the understanding of the region of interest (ROI).

Consider deploying $N$ UAVs that communicate in a noisy environment and independently plan their respective trajectories. Each UAV $i$ maintains a local map belief $\mathcal{M}_i$, which represents its current understanding of the ROI. The planned trajectory is represented as a sequence $\psi_i = (p_0, p_1, \dots, p_B)$, where $p \in \mathbb{R}^3$ denotes the UAV's sampling position over the ROI, and $B \in \mathbb{N}^+$ represents the communication and sampling budget. The UAVs aim to maximize their understanding of the terrain, i.e., reduce belief uncertainty. According to Westheider et al. \cite{RL-IPP-2-baseline}, the information gained from sampling along a trajectory can be quantified using an information-theoretic criterion:
\begin{equation}
    \mathrm{I}(\psi) = \sum_{j=0}^{B-1} \big[ H(\mathcal{M} \mid z_{j}, \boldsymbol{p}_{j}) - H(\mathcal{M} \mid z_{j+1}, \boldsymbol{p}_{j+1}) \big],
\end{equation}
where $H$ denotes the Shannon entropy of the belief map. Thus, the final training objective is defined as follows:
\begin{equation}
\psi^* = \arg\max_{\psi \in \Psi^N} I(\psi), \quad \text{s.t.} \ \forall \ i, \ \text{Cost}(\psi_i) \leq B.
\end{equation}
This formula is a generalized description of the IPP problem.

We use the MARL method to solve this problem and define the key parameters as follows.

 {\bf Global environment state}: $s^t = \{ \mathcal{M}, p^t_{1:N}, b \} \in S$, representing the global belief map, the positions of the $N$ UAVs, and the task budget.

 {\bf Action space}: $U = \{x^+, x^-, y^+, y^-, z^+, z^-\}$, representing the 6 directions along the coordinate axes.

 {\bf Reward}: $R^{t}\left(s^{t}, \boldsymbol{u}^{t}, s^{t+1}\right) = \alpha \frac{H\left(\mathcal{M}^{t+1}\right) - H\left(\mathcal{M}^{t}\right)}{H\left(\mathcal{M}^{t}\right)} + \beta,$
representing the relative reduction in map entropy, which is our objective. $\alpha$ and $\beta$ denote scaling factors.

At each time step, the actions of all UAVs, i.e., their next sampling locations, are defined as a joint action $u^t = \{ u_1^t, \dots, u_N^t \} \in U^N$. The global reward is then distributed to individual UAVs using the COMA algorithm \cite{COMA}, which is a centralized learning, decentralized execution method detailed in Section \ref{method:ACnet-COMA}.

\section{Methodology}
\label{method}
In this section, we introduce our novel MARL-based UAV path planning framework. Subsection \ref{method:overall-archi} provides an overview of the framework, followed by Subsections \ref{method:actor-critic} and \ref{method:overall-archi}, elaborating on its components. Finally, Subsection \ref{method:ACnet-COMA} details the training process.
\subsection{Overall Architecture}
\label{method:overall-archi}
\begin{figure*}[ht]
    \centering
    \includegraphics[width=\linewidth]{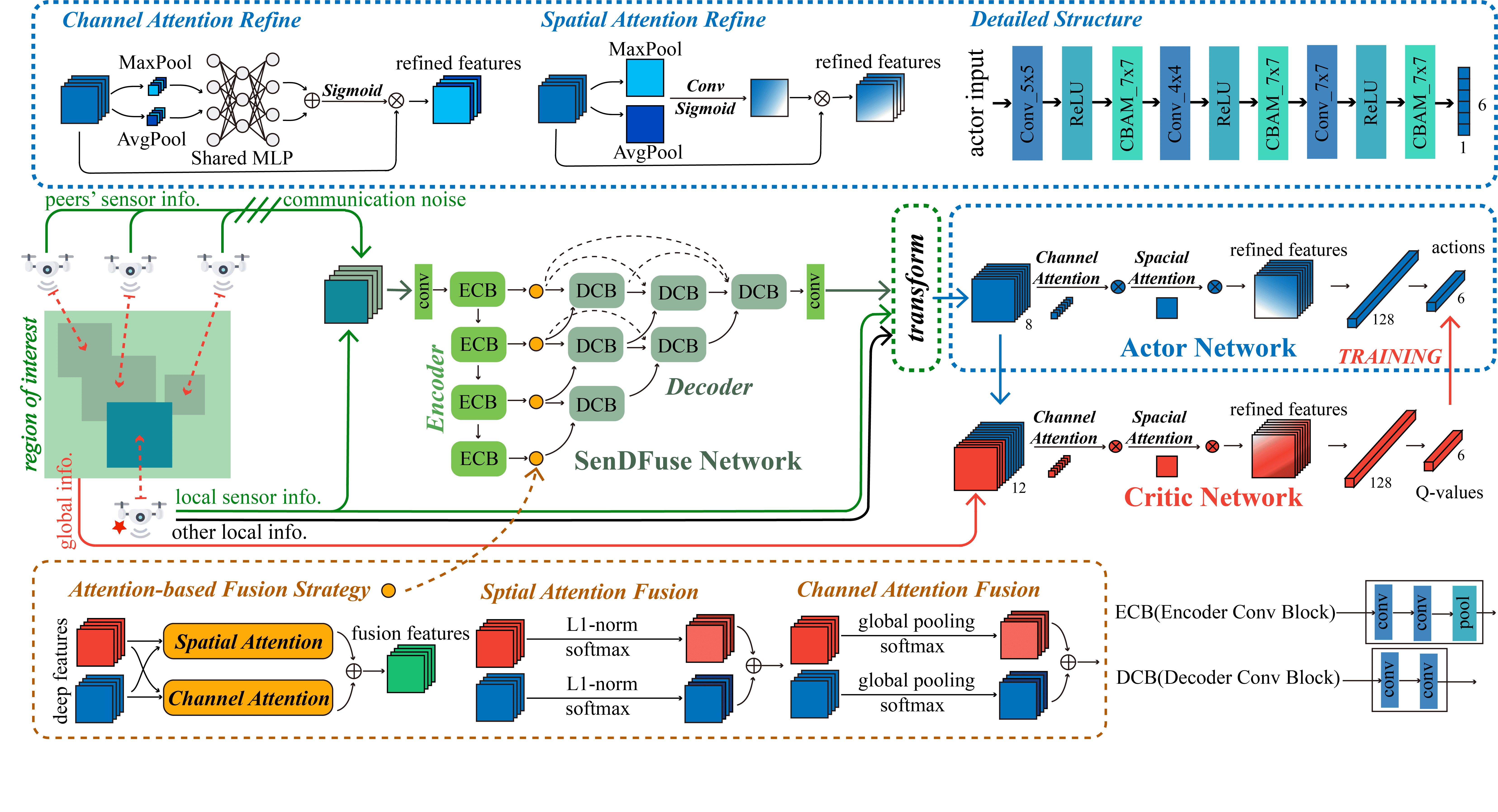}
    \caption{The overall structure of our proposed framework primarily illustrates the data flow and processing modules. Multiple drones sample an area of interest, and each drone fuses the visual data from multiple agents using the SenDFuse Network. The fused visual perspective is then combined with other local information and stacked as input to the Actor network, which makes decisions. The input to the Critic network includes both the input to the Actor network and global information, serving to evaluate the performance of the Actor network.}
    \label{fig:overallarchi2}
\end{figure*}

In this subsection, we present the overall structure of our proposed framework illustrated in Figure \ref{fig:overallarchi2}, where multiple UAVs collaboratively perform sensor sampling over an ROI. These UAVs communicate within a constrained communication and sampling budget in a noisy environment. The figure highlights the UAV marked with a star, which receives noisy communication data from three other UAVs. The received data is processed by the SenDFuse Network (i.e., Sensor Denoising and Fusion Network). Note that besides sensor data, drones also exchange position and altitude information as well as identifiers. Before performing SenDFuse Net, the sensor data needs to be aligned.

All UAVs share a unified Actor network for decision-making. Multi-sensor data, after being denoised and fused by the SenDFuse Network, is combined with local information detailed. These inputs are preprocessed and transformed into an 8-channel input for the Actor network. The Actor network adopts a Convolutional Block Attention Module (CBAM)  structure \cite{CBAM}, which processes the input into a probability distribution over a 6-dimensional discrete action space to guide the UAVs decision-making at each step. The Critic network evaluates the actions taken by the Actor network. While sharing a similar CBAM structure with the Actor network, it additionally incorporates four dimensions of global information. 

\subsection{SenDFuse Network}
\label{method:sendfuse}
In this subsection, we introduce the SenDFuse Network (i.e., Sensor Denoising and Fusion Network) illustrated in the lower part of the Figure \ref{fig:overallarchi2}. This network is used to denoise and fuse the sensor information of $n$ UAVs, generating a globally fused vision map.
We extend the NestFusion method \cite{NestFusion} from fusing two pieces of information with the same mode but different modalities (e.g., visible light and infrared images taken at the same location) to fusing $n$ pieces of information with the same modality but different modes (e.g., images of the same type taken at different locations). Additionally, we fully exploit the noise resistance capability of NestFusion as an autoencoder-like structure and integrate it into the framework. 


\subsubsection{Fusion Strategy}
We propose an attention-based fusion strategy to process inputs from multiple sensors. Let $\Phi$ represent the encoded sensor data set from $n$ UAVs, where the sensor data $s$ includes communication noise $\epsilon$ and $\phi(\cdot)$ denotes the encoder of the network. Note that UAV 0 represents the local UAV, and its data is unaffected by noise. The fusion strategy $\mathcal{F}(\cdot)$ can be expressed as:
\begin{equation}
\Phi=\big \{ \phi(s_0), \phi(s_1+\epsilon_1), ..., \phi(s_{n-1}+\epsilon_{n-1})\big \}.
\end{equation}
$\mathcal{F}(\cdot)$ computes a weighted sum of spatial and channel attention over $\Phi$:
\begin{equation}
\mathcal{F} ( \Phi ) = \alpha \ \mathcal{C}(\Phi)+\beta \ \mathcal{S}(\Phi),
\end{equation}
where $\mathcal{C}(\cdot)$ represents channel attention, $\mathcal{S}(\cdot)$ represents spatial attention and $\alpha + \beta = 1$.
Given that $\Phi$ has a shape of $[d_1, d_2, ..., d_m]$, for any given channel $c$, the attention formulas are as follows:
\\Channel Attention:
\begin{equation}
\mathcal{C}(\Phi)[c, *] = \frac{\exp\big(\text{Pooling}(\Phi[c])\big)}{\sum_{i=1}^m \exp\big(\text{Pooling}(\Phi[i])\big)} \cdot \Phi[c, *],
\end{equation}
Spatial Attention:
\begin{equation}
\mathcal{S}(\Phi)[c, *] = \frac{\exp\big(\|\Phi[:, *]\|_1\big)}{\sum_{*} \exp\big(\|\Phi[:, *]\|_1\big)} \cdot \Phi[c, *].
\end{equation}
Here, $\|\Phi[:, *]\|_1$ represents the L1 norm of the channel vector at a fixed spatial position, and $\sum_*$ denotes summation over all spatial dimensions. 
Let $R$ represent the fused image and let $\Phi$ denotes the decoder. Then, the extended fusion and denoising process can be defined as:
\begin{equation}
R = \psi \big( \mathcal{F} ( \Phi ) \big).
\end{equation}
Notably, in the spatial attention formula, $c$ does not directly participate in the computation but is used to index the output for each channel.



\subsubsection{Training \& Deployment for SenDFuse}
\label{method:fusenet:train-deploy}
The SenDFuse Network is pre-trained offline by generated data and then deployed within subsequent architectures. During training, the fusion strategy is temporarily disabled, and the network is trained as a one-to-one AE. Specifically, the network is tasked with encoding and reconstructing single images to establish its capability for image representation and reconstruction. 

\begin{figure}[t]
    \centering
    \includegraphics[width=\linewidth]{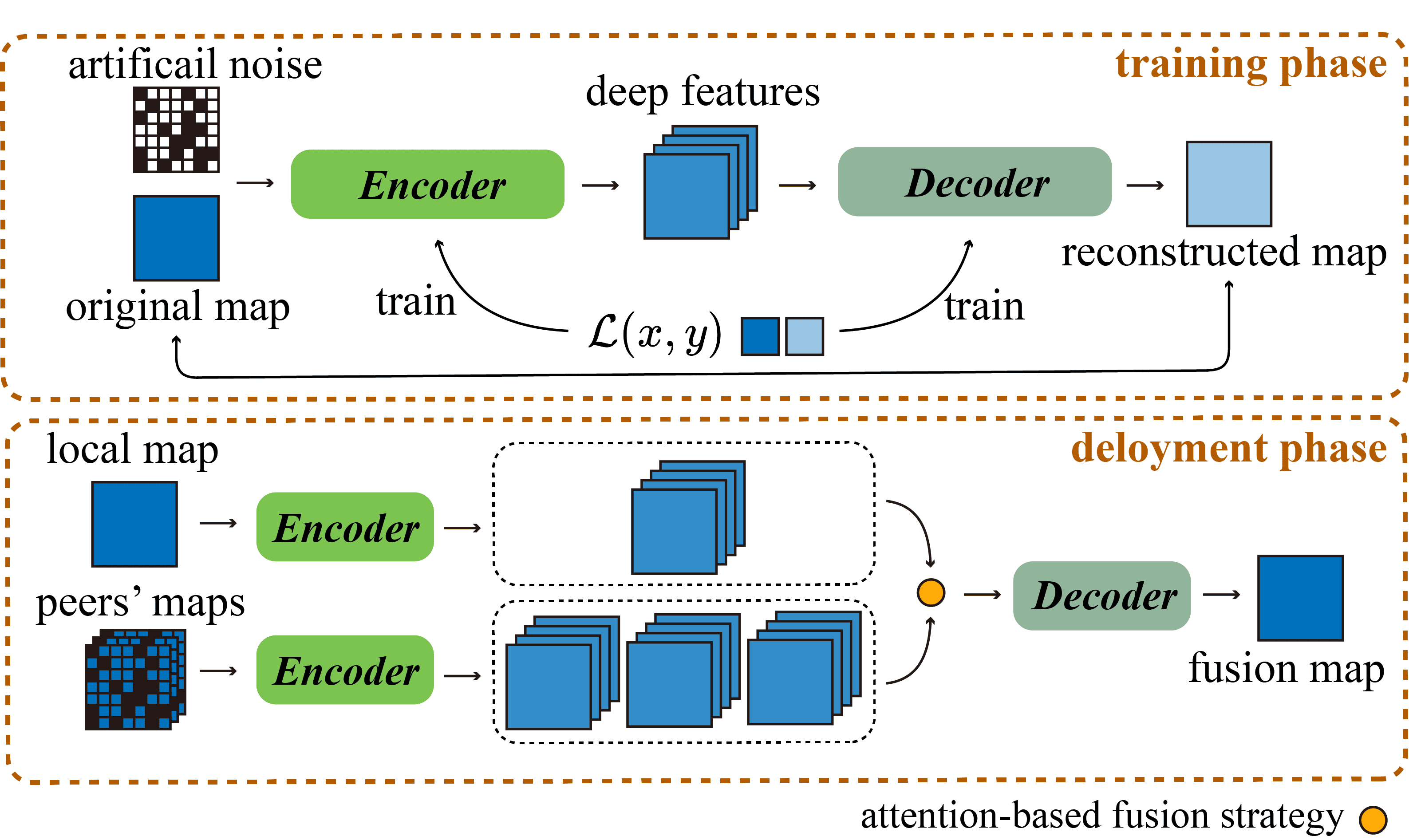}
    \caption{The SenDFuse Network operates in two phases: training and deployment. During training, artificial noise is added to the input data, and the network is trained to reconstruct noise-free images. In the deployment phase, an attention-based fusion strategy is applied to process deep-level features. }
    \label{fig:tain-and-deploy}
\end{figure}

To address the noise in input images, we adopt the methodology of denoising autoencoder \cite{DenoisingAE}. During training, artificial noise according to equation \ref{eq:noise} is randomly added to the input images. The network is then trained to regress the output images toward the clean, noise-free versions, as illustrated in Figure \ref{fig:tain-and-deploy}. This two-phased procedure enhances the network's robustness to noisy inputs. 

The loss function used in SenDFuse Network is a linear combination of pixel-level loss and structural loss. The pixel-level loss leverages both the Mean Squared Error (MSE) loss and the Mean Absolute Error (MAE) loss, while the structural loss employs the Structural Similarity Index Measure (SSIM) loss. SSIM is a perceptual metric that quantifies the structural similarity between two images, with values ranging from $[-1, 1]$, where a value closer to $1$ indicates higher similarity between the images. The loss function is defined as:
\begin{equation}
\label{eq:sendfuse-losee}
\mathcal{L}(x, y) = \alpha \text{MSE}(x, y) + \beta \text{MAE}(x, y) + \gamma (1 - \text{SSIM}(x, y)).
\end{equation}
Here, $\alpha$, $\beta$ and $\gamma$ are scaling parameters used to balance the magnitude of the three loss components.

\subsection{Actor \& Critic Network}
\label{method:actor-critic}
\label{method:ACnet}
In this section, we introduce our Actor \& Critic networks. In RL, the Actor is responsible for learning and executing the policy, selecting actions based on the current state, and the Critic evaluates the quality of the actions chosen by the Actor. CBAM \cite{CBAM} is a convolutional neural network module designed to enhance feature representation by combining channel attention and spatial attention. It first uses the channel attention module to adaptively select important feature channels, followed by the spatial attention module to focus on critical feature locations. The structure of the Actor network is shown in the upper part of figure \ref{fig:overallarchi2}, while the Critic network has a similar architecture, differing only in the number of input channels.


It is worth noting that the spatial-channel attention mechanism in the CBAM module differs slightly from the fusion strategy used in the SenDFuse Network. Specifically, the channel attention in CBAM includes an additional shared, trainable MLP structure. For spatial attention, the SenDFuse Network typically employs a single pooling strategy, whereas the CBAM module combines two pooling strategies.

After obtaining local information, the drone will process it to generate eight $11 \times 11 $ submaps. These submaps will serve as the input to the Actor network. 
(a) Remaining Budget Map: Each element represents the ratio of the remaining budget to the total budget. 
(b) Drone Identifier Map: Each element represents the ratio of the current drone's ID to the total number of drones. 
(c) Altitude Map: The altitude of all drones as a ratio to the total altitude. If a location has no drone, the value is 0. 
(d) Footprint Map: The positions visited by the drone on the map. 
(e) Local Sensor Observation Map (map-sensor-local): The results of local sensor sampling.  
(f) Local Map Belief: $\mathcal{M}_i$ from \ref{problem-statement}, which represents the occupancy probability map. 
(g) Shannon Entropy of Local Belief Map: $H(\mathcal{M}_i \mid z, p)$.
(e) Denoising-Fusion Map: The denoising and fusion map generalized by SenDFuse Network. 

The input to the Critic network, in addition to all the aforementioned submaps, also includes four global information maps:
(a) Global Map Belief: The global map belief $\mathcal{M}_i$. 
(b) Shannon Entropy of Global Map Belief: $H(\mathcal{M} \mid z, p)$. 
(c) Global Footprint Map: The map that marks all visited positions globally. 
(d) Actions of All Drones: This is crucial for the COMA algorithm introduced in Section \ref{method:ACnet-COMA}.

By designing the network input in this way, the Actor-Critic network can achieve proper collaboration. We train the networks to process the input data from each channel and obtain appropriate results.

\subsection{Training Algorithm}
\label{method:ACnet-COMA}
In this subsection, we introduce the train algorithm for our framework. COMA \cite{COMA, RL-IPP-2-baseline} is a centralized training and decentralized execution algorithm for MARL. It computes counterfactual baselines through a centralized value function to evaluate the advantage of each agent's actions, reducing the non-stationarity and signal mixing issues among agents. This enables efficient policy optimization and is suitable for cooperative scenarios.

The Critic network evaluates the policy $\pi$ by estimating the state-action value function $Q_t((s^t, u^t))$, where $u^t = (u_1^t, ...,  u_n^t)$, and the actions of other agents, $u_{-i}^t$, are fixed. The Critic network is trained using $TD(\lambda)$ function to estimate the discounted return $G_t$. During training, it uses the global state $s^t$, while the Actor network only uses local information $\omega_i^t$, to predict decentralized actions.

To measure the contribution of  action $u_i^t$ taken by agent $i$ to the team, a counterfactual baseline is used to calculate the advantage function $A_i^t(s^t, u^t)$:
\begin{equation}
\label{eq:advantage}
A_i^t(s^t, u^t) = Q_t(s^t, u^t) - \sum_{u_i'^t \in U} \pi(u_i'^t | \omega_i^t) Q_t(s^t, (u_{-i}^t, u_i'^t)).
\end{equation}
When optimizing the policy $\pi(\cdot \mid \omega_i^t)$, the policy gradient theorem and Equation \ref{eq:loss} are used to minimize the following objective:
\begin{equation}
\label{eq:loss}
\mathcal{L} = -log \big(\pi(u_i^t|\omega_i^t) A_i^t(s^t, u^t)\big).
\end{equation}
We present the COMA pseudocode in Algorithm \ref{coma-code} to describe the training process for our mission. Through the COMA algorithm, the system can reasonably allocate the global reward to each agent.

\section{Experiment}
\label{experiment}

In this section, we present our experimental results. The proposed framework demonstrates superior performance in both the training phase of MARL and the deployment phase, particularly in noisy environments during multi-UAV scanning tasks over ROI. 

\begin{figure}[ht]
    \centering
    \includegraphics[width=\linewidth]{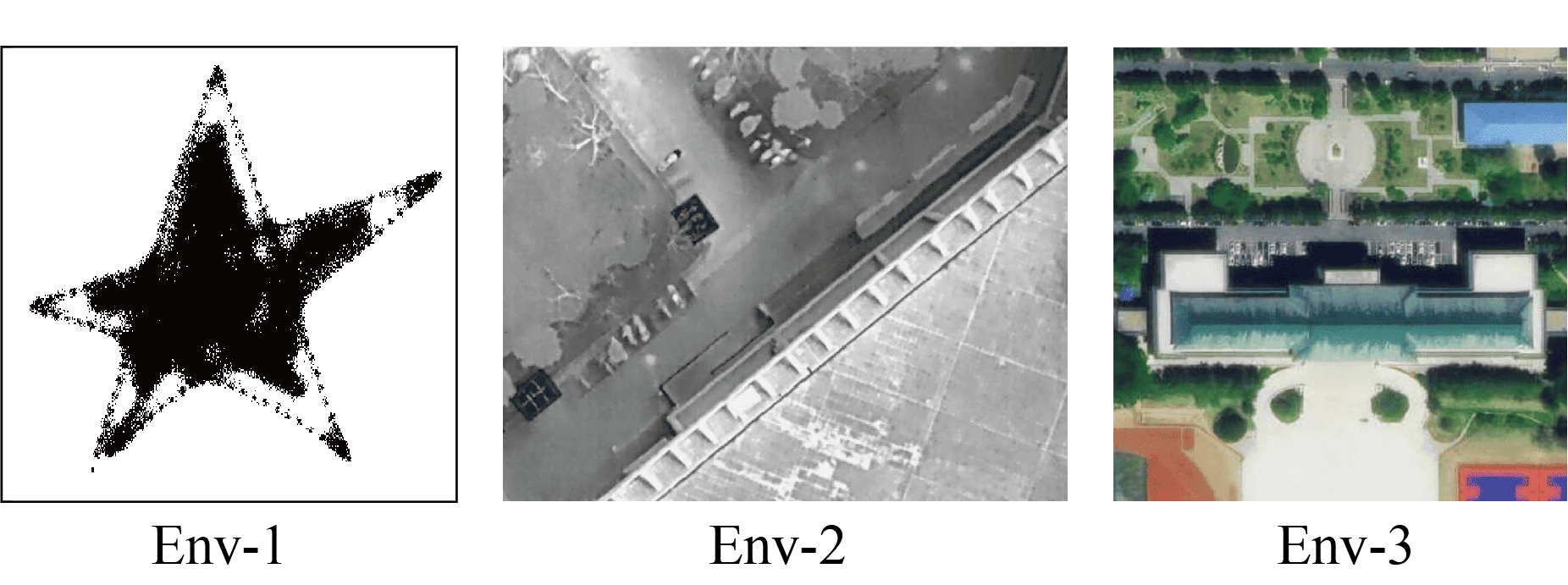}
    \caption{Our experimental environments includes three types: Env1: synthetic data, Env2: thermal imaging data, and Env3: visible-light data. Note that in Env3 we stipulate that drones are not allowed to fly into building areas, and the sampling category for building areas is designated as valueless.}
    \label{fig:env123}
\end{figure}

\subsection{Set-up}
\label{exp:exp-set-up}

We conducted experiments in three distinct environmental categories: 
(1) Env1: A manually drawn star-shaped region with a certain degree of random structure, where black pixels are defined as valuable areas.
(2) Env2: A thermal imaging UAV image from the HIT-UAV dataset \cite{hit-uav}. The UAV is equipped with a FLIR long-wave infrared thermal camera in white-hot mode. We define areas with relatively high surface temperatures as valuable regions.
(3) Env3: An aerial view of Geology Palace at Chaoyang Campus, Jilin University, captured using a conventional visible-light camera. We define vegetation areas as valuable regions, meaning the green areas excluding rooftops are considered valuable.
All environments are downsampled to appropriate resolutions to allow uniform partitioning into rectangular ROIs and The total number of sampled pixels should exceed 250,000 to ensure experimental validity. A homogeneous fleet of drones is deployed in each environment for RL training, and the trained Actor networks are extracted for deployment and testing. Although it is stipulated that the drone will not fly out of the area of interest, it may still sample content from undefined regions. In this case, we define the sampling result as 0, representing a valueless category. For all experiments, we conducted 10 trials and computed the mean and standard deviation for visualization and tabulation.

\begin{algorithm}[t]
    \caption{COMA Training}
    \label{coma-code}
    \begin{algorithmic}[1]
    \REQUIRE{ROI $P$, UAV count $N$, actor $\pi_\theta$, critic $Q_\phi$, learning rates $\alpha_\pi$, $\alpha_Q$, discount factor $\gamma$, trace decay $\lambda$, batch size $B$,global map belief $\mathcal{M}$, budget $bgt$, SenDFuse Network $\psi$, $\phi$ and $\mathcal{F}$}
    \STATE Initialize $\pi_\theta$, $Q_\phi$ and $\mathcal{M}$, replay buffer $\mathcal{B} = \emptyset$
    \FOR{each episode}
        \STATE Reset environment, initial state $s^0 = \{\mathcal{M}^0, p^0_{1:N}, b^0\}$
        \FOR{$t = 0$ to $bgt - 1$}
            \FOR{each UAV $i$}
                \STATE Perform SenDFuse $R = \psi \big( \mathcal{F} ( \Phi ) \big)$
                \COMMENT {See \ref{method:sendfuse}}
                \STATE Build observation $\omega_i^t$ 
                \COMMENT{See \ref{method:actor-critic}}
                \STATE Sample action $u_i^t \sim \pi_\theta(u_i^t | \omega_i^t)$
            \ENDFOR
            \STATE Execute $\mathbf{u}^t = \{u_1^t, \dots, u_N^t\}$ and update $\mathcal{M}$
            \STATE Get $s^{t+1}$ and compute reward $r^t$
            \STATE Store transition $(s^t, \mathbf{u}^t, r^t, s^{t+1})$ in $\mathcal{B}$
            \IF{$|\mathcal{B}| \geq B$}
                \STATE Sample batch of transitions from $\mathcal{B}$
                \STATE Initialize $G = 0$ \COMMENT{For $TD(\lambda)$ calculation}
                \FOR{each transition $(s, \mathbf{u}, r, s')$ in reversed batch order}
                    \STATE Compute $G = r + \gamma G$ \COMMENT{Recursive $TD(\lambda)$}
                \ENDFOR
                \STATE Update critic: $L_{\text{critic}} = \frac{1}{|B|} \sum (Q_\phi(s, \mathbf{u}) - G)^2$
                \FOR{each UAV $i$}
                
                    
                    \STATE  Compute advantage:  

                    \vspace{-15pt}
                    \begin{align*}
                    b_{cf} &= \sum_{u_i'^t} \pi_\theta(u_i'^t | \omega_i^t) 
                    Q_\phi(s^t, (u_{-i}^t, u_i'^t)) \\
                    A_i^t &= Q_\phi(s^t, \mathbf{u}^t) - b_{cf}
                    \end{align*}
                    \vspace{-15pt}
                    \STATE Update actor: $L_{\text{actor}} = - \mathbb{E}_{\pi_\theta} [\log \pi_\theta(u_i^t | \omega_i^t) A_i^t]$
                \ENDFOR
                \STATE Perform gradient updates for $\theta$ and $\phi$ using $\alpha_\pi$ and $\alpha_Q$
                \STATE Clear buffer $\mathcal{B}$
            \ENDIF
        \ENDFOR
    \ENDFOR
    \end{algorithmic}
    \end{algorithm}


We model the noise as a combination of distance-dependent multiplicative attenuation and additive Gaussian white noise, which is one of the most representative types of noise in nature \cite{gauss-noise, multi-noise}. Let $I(x,y)$ denote the original image channel, then the received image is given by
\begin{equation}
\label{eq:noise}
    I_r(x, y) = \alpha(x,y) I(x, y) + n(x, y)
\end{equation}
where $x, y$ are the coordinates, $\alpha(x,y)$ is the distance-dependent attenuation factor, and $n(x, y) \sim \mathcal{N}(0, \sigma^2)$ represents Gaussian white noise. 
We define the moderate noise as $\alpha \sim \mathcal{U}(0.8, 1)$ and $\sigma=0.02$, while loud noise is characterized by $\alpha \sim \mathcal{U}(0.6, 1)$ and $\sigma=0.06$. These definitions will appear in the following text.

\subsection{MARL Training}
\label{exp:training}

Our training process follows the algorithm \ref{coma-code} under the scenario outlined in \ref{exp:exp-set-up} and we present results in Env3. During training, we deploy 4 UAVs, with each mission starting from fixed initial positions and operating under a communication budget of $b = 15$. The total number of episodes is set to 1200, and the Actor and Critic networks are updated using the Adam optimizer, where the learning rate for the Actor is $1e\text{-}5$, for the Critic is $1e\text{-}4$, and the batch size is 512. We set $\lambda = 0.8$ and the discount factor $\gamma = 0.99$ for $TD(\lambda)$. We let $\alpha = 10$ and $\beta = -0.17$ in reward function in Section \ref{problem-statement}.

We highlight the progression of the global reward defined in \ref{problem-statement} as training episodes advance and use this to analyze the effectiveness of the methods. In fully cooperative tasks such as UAV Swarming, the global reward effectively reflects the progress of system training. 

\begin{figure}[ht]
    \centering
    \includegraphics[width=\linewidth]{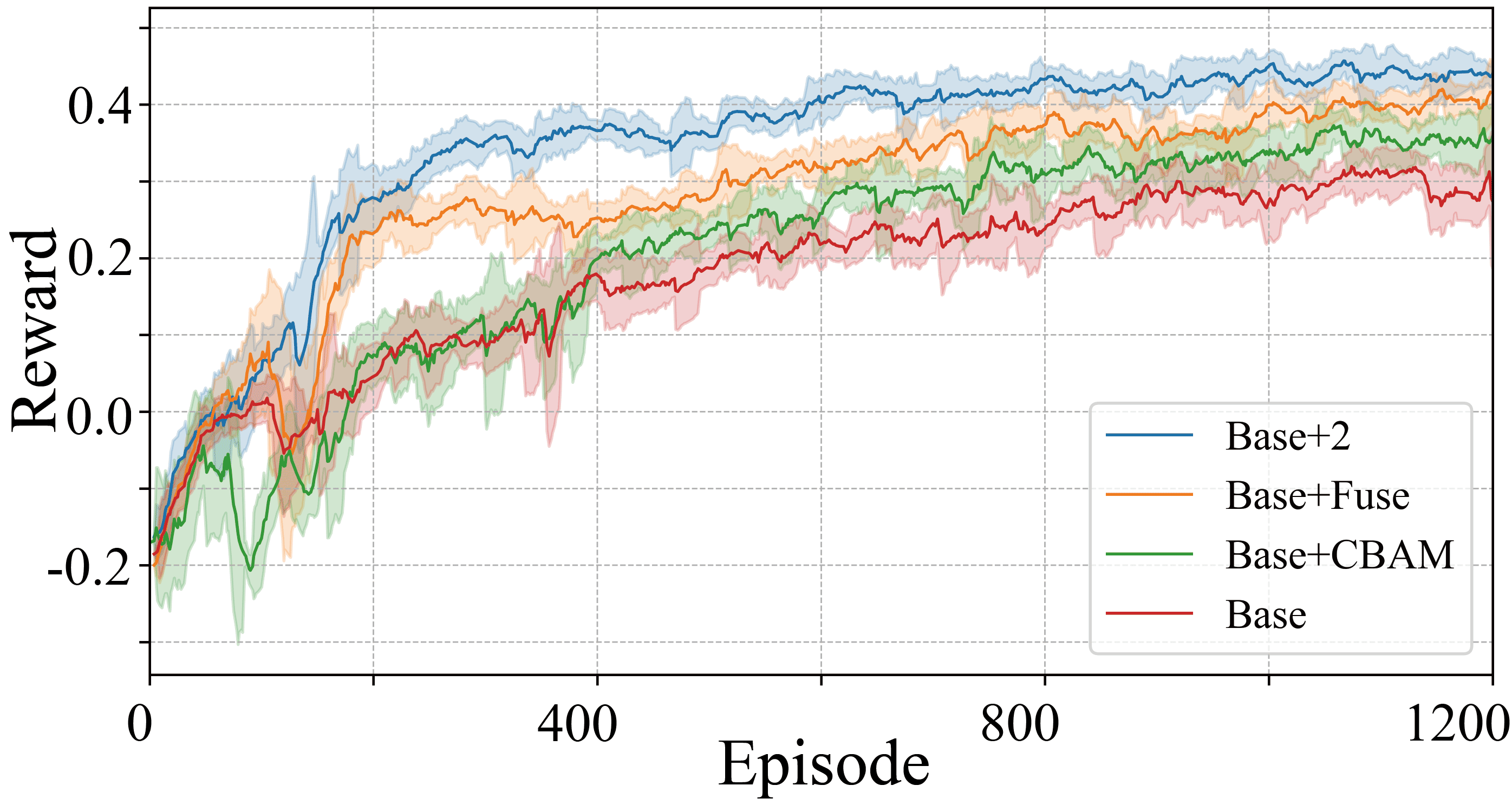}
    \caption{Env3: Reward variations for each method over 1,200 training episodes. Among them, Base represents method with the absence of SenDFuse and CBAM. The method incorporating both major modules performs the best in terms of both convergence speed and final convergence value.}
    \label{fig:reward}
\end{figure}

Figure \ref{fig:reward} shows how the reward varies over the course of training in Env3 . As training progresses, all methods eventually converge. However, the model incorporating both CBAM and SenDFuse structures exhibits the best performance in terms of convergence speed and final convergence value. The effect of using Fusion alone is superior to that of using CBAM alone. This is because our training environment simulates moderate noise, and the network designed for fusion and denoising demonstrates its effectiveness, particularly in the early stages of training (around 150-200 episodes). It is also worth noting that although CBAM does not exhibit the same noise-resistant capability as SenDFuse, it still performs better than the naive model. This could be attributed to the fact that the channel-spatial attention mechanism learns certain denoising and fusion capabilities through the RL learning process. This hypothesis is further supported by subsequent experiments.

\begin{figure}[t]
    \centering
    \includegraphics[width=\linewidth]{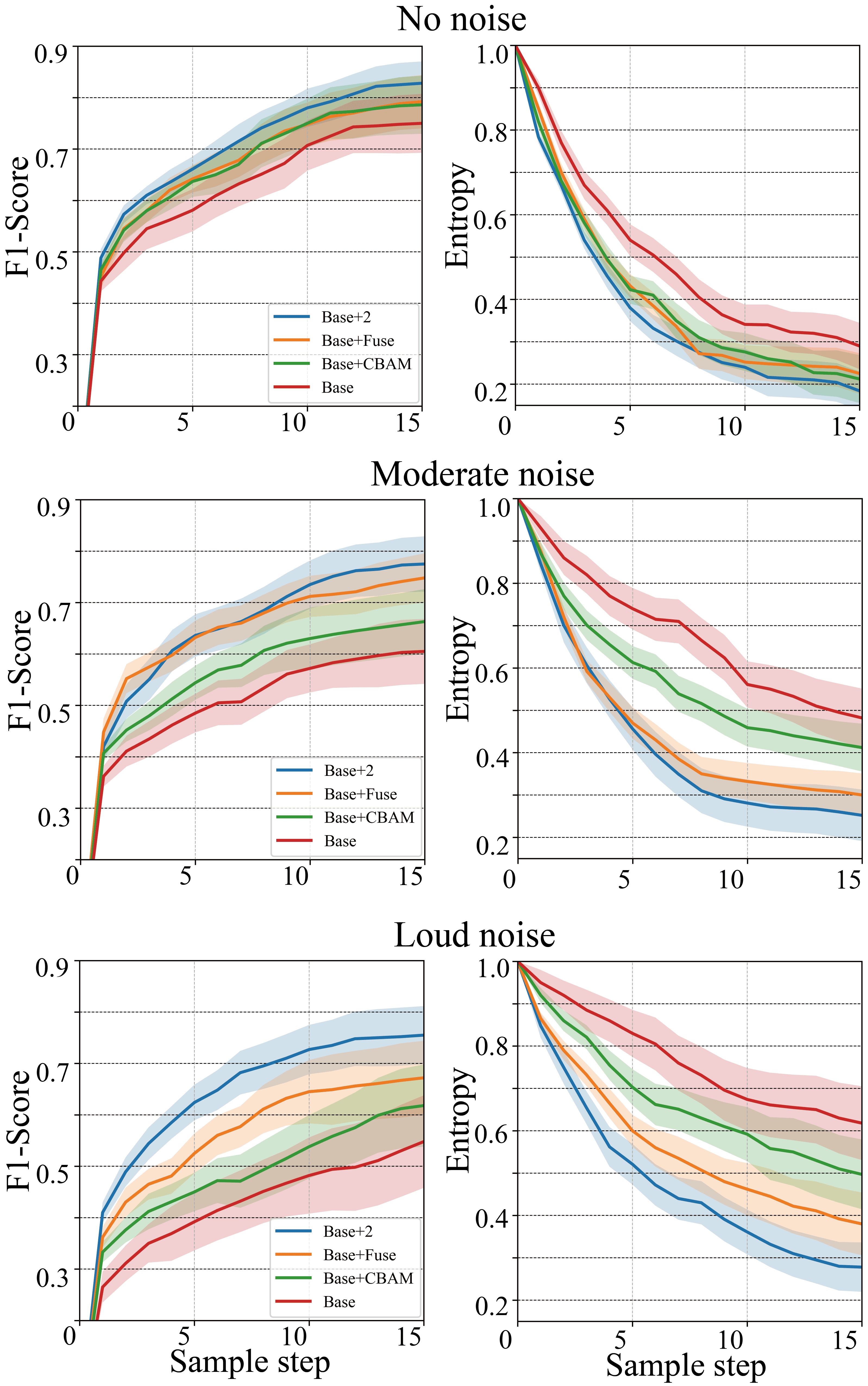}
    \caption{Env3: The variations of F1-Score and Entropy with respect to the steps of sampling. As the noise intensity increases, the importance of the two main modules becomes more evident.}
    \label{fig:EF-1}
\end{figure}

\subsection{Deployment}
\label{exp:deployment}

\subsubsection{Ablation Study}
We deploy the trained and converged models into the environment shown in Env3 for deployment testing to evaluate their performance. We select F1-Score and shannon entropy of the system's global map belief defined in \ref{problem-statement} to measure the performance during deployment testing. Since the task is essentially a binary classification problem determining whether each pixel in the ROI is valuable or not, we also focus on the F1-Score to assess task performance. Entropy represents the confidence level of the system's global map belief, reflecting the system's understanding and certainty regarding the ROI. In the Env3 environment, we simulate three different noise conditions: no noise, moderate noise, and loud noise, to evaluate the module's performance under varying levels of uncertainty.

\begin{table*}[t]
    \centering
    \caption{Performance Comparison Across Different Environments}
    \label{table:1}
    \renewcommand{\arraystretch}{1.2}
    \begin{tabular}{lc ccc ccc}
        \toprule
        & & \multicolumn{3}{c}{F1-Score $\uparrow $} & \multicolumn{3}{c}{Entropy $\downarrow$} \\
        \cmidrule(lr){3-5} \cmidrule(lr){6-8}
        Env & Method & 33\% & 67\% & 100\% & 33\% & 67\% & 100\% \\
        \midrule
        \multirow{4}{*}{Env1} & Ours   & $\textbf{0.5509} \pm 0.0127$ & $\textbf{0.7807} \pm 0.0252$ & $\textbf{0.8432} \pm 0.0260$ & $0.7216 \pm 0.0233$ & $\textbf{0.3883} \pm 0.0491$ & $\textbf{0.2121} \pm 0.0508$ \\
                              & AG      & $0.4631 \pm 0.0512$ & $0.6242 \pm 0.0763$ & $0.6734 \pm 0.0250$ & $\textbf{0.7119} \pm 0.0455$ & $0.4273 \pm 0.0572$ & $0.2865 \pm 0.0579$ \\
                              & NL     & $0.3084 \pm 0.0916$ & $0.5211 \pm 0.0423$ & $0.6521 \pm 0.0455$ & $0.7538 \pm 0.0731$ & $0.4528 \pm 0.0514$ & $0.3699 \pm 0.0527$ \\
                              & Random & $0.3271 \pm 0.1752$ & $0.4638 \pm 0.1809$ & $0.5125 \pm 0.1820$ & $0.8876 \pm 0.0404$ & $0.6852 \pm 0.0842$ & $0.5867 \pm 0.1121$ \\
        \midrule
        \multirow{4}{*}{Env2} & Ours   & $\textbf{0.5206} \pm 0.0205$ & $\textbf{0.7357} \pm 0.0334$ & $\textbf{0.8508} \pm 0.0341$ & $\textbf{0.7163} \pm 0.0195$ & $\textbf{0.3597} \pm 0.0375$ & $\textbf{0.2285} \pm 0.0384$ \\
                              & AG     & $0.3323 \pm 0.0588$ & $0.6843 \pm 0.0632$ & $0.8471 \pm 0.0725$ & $0.8148 \pm 0.0410$ & $0.5872 \pm 0.0508$ & $0.4154 \pm 0.0563$ \\
                              & NL     & $0.2931 \pm 0.1204$ & $0.5628 \pm 0.0738$ & $0.6225 \pm 0.0565$ & $0.8691 \pm 0.0883$ & $0.6247 \pm 0.0435$ & $0.5173 \pm 0.0411$ \\
                              & Random & $0.2016 \pm 0.1545$ & $0.4279 \pm 0.1920$ & $0.5211 \pm 0.1937$ & $0.8523 \pm 0.0381$ & $0.6522 \pm 0.0798$ & $0.5235 \pm 0.0997$ \\
        \bottomrule
    \end{tabular}
\end{table*}

The analysis results indicate that in the no noise situation, except for the naive method (Base), which performs the worst in both F1-Score and Entropy, the other three methods show little difference in performance. Overall, the method incorporating both modules performs the best, while the methods with only one module perform similarly, slightly inferior to the one with both.
In the moderate noise situation, the impact of the modules becomes more apparent. The model with both modules continues to dominate, followed closely by the model with only the Fusion module. However, the model with only the CBAM module falls further behind but still outperforms the naive method.
In the loud noise situation, the performance gap between the models becomes even more pronounced, further validating our analysis. That is, in noisy environments, the Fusion module serves as an effective sensor denoising and fusion method, while the CBAM module, as an attention mechanism, can also learn certain denoising and fusion capabilities during the RL process. Although the CBAM module alone is not as effective as the Fusion module alone, it still performs better than the naive method. It is also worth noting that as noise increases, all methods exhibit instability (with increasing variance). However, the Fusion module helps mitigate this trend to some extent.
Ultimately, these two modules can complement each other, and incorporating both achieves the best overall performance.

\subsubsection{Comparison against Other Approaches}
We deploy other IPP methods in the Env3 environment and compare them with our method. These methods include: 
(1) ``AG'' (Adaptive Gain): An adaptive IPP method proposed by Carbone et al. \cite{iros-2}. This is a monitoring and mapping strategy that adaptively selects target areas based on expected information gain, which measures the potential for uncertainty reduction through further observations. 
(2)  ``NL'' (Non-adaptive Lawnmower): A non-adaptive lawnmower-like method that deploys UAVs uniformly at equal intervals to perform straight-line scanning.
(3) ``Random'': UAVs make completely random decisions.
To simulate a more generalized environment, the noise intensity is set to moderate. Note that for NL and random, which are non-communication-based algorithms, environmental noise does not affect their performance.

\begin{figure}[h]
    \centering
    \includegraphics[width=\linewidth]{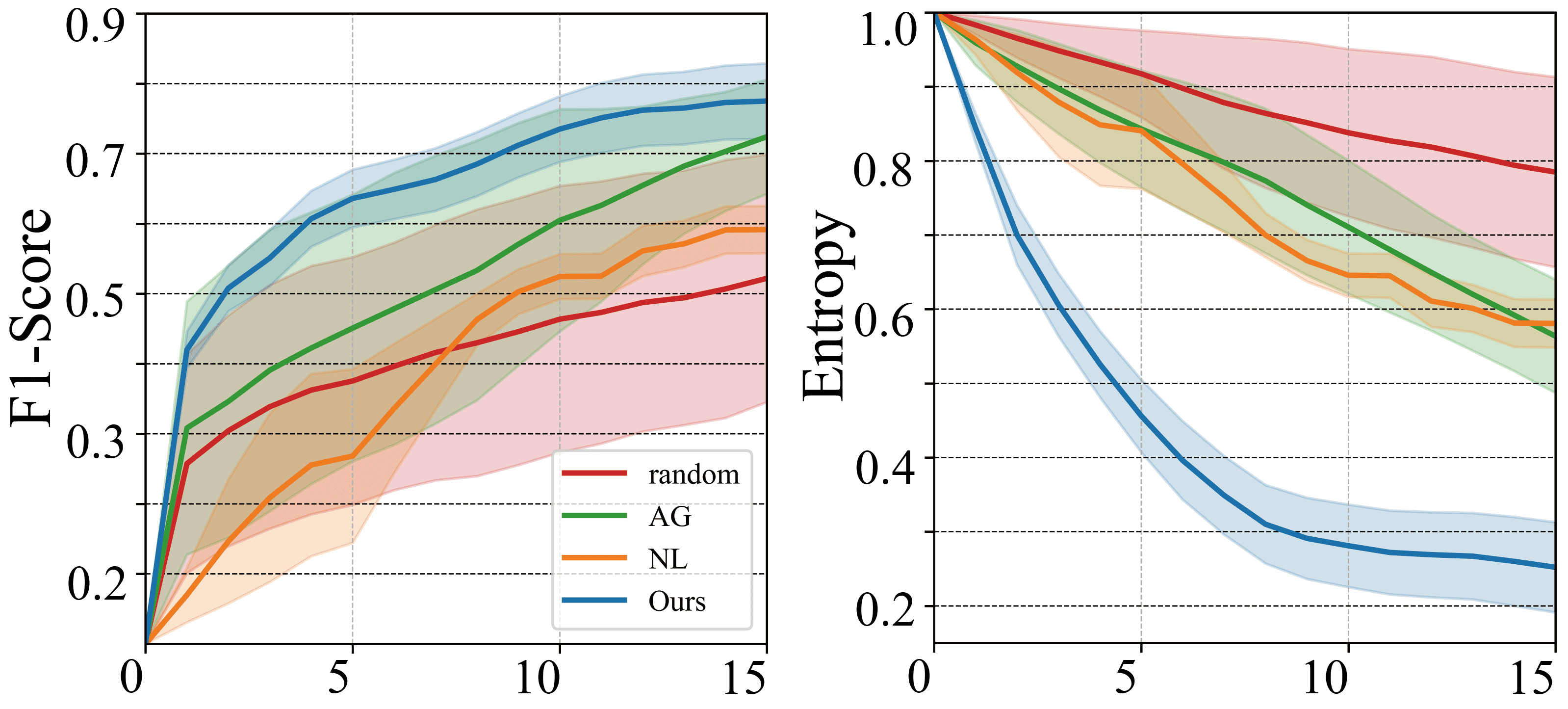}
    \caption{Env3: Comparison of the deployment test results of our method between other 3 methods in the same scenario. The two adaptive methods outperform the two non-adaptive methods, with our method achieving the best overall performance.}
    \label{fig:comp}
\end{figure}

The experiments demonstrate the superiority of our method compared to the three baselines, achieving the best performance and stability in both F1-Score and Entropy metrics.
Notably, the AG method also performs well in terms of F1-Score but performs worse in Entropy. 
We speculate that this is because AG adopts a relatively conservative strategy: it tends to make clear predictions about the explored areas, which helps with accurate classification and benefits the F1-Score. However, this also leaves many unexplored regions, causing the entropy to decrease slowly.

We conduct the same experiments on Env1 (synthetic data) and Env2 (thermal imaging data) and compare the deployment performance of the four methods. We keep the training parameters unchanged and trained the model until convergence, then fixed the noise intensity to a moderate level. The experimental results including mean and standard deviation of F1-score and entropy are shown in Table \ref{table:1}. Experiments show that our method demonstrates superiority across various ROI structures. Best results are highlighted in bold.

\section{Conclusion}
\label{conclusion}

In our paper, we propose a novel MARL framework with a robust communication protocol and composite attention mechanism to solve the problem of multi-UAV cooperative information collection in a full-3D environment.
We develop the SenDFuse Network for multi-sensor denoising and fusion, and employ an attention-enhanced Actor-Critic network, significantly improving multi-UAV collaboration efficiency, especially in noisy environments. We utilize the COMA algorithm for credit assignment within the framework and demonstrate that it outperforms existing algorithms in both the training and deployment phases.

\bibliographystyle{IEEEtran}
\bibliography{references}
\end{document}